\documentclass[11pt]{article}  
\usepackage{latexsym}
\usepackage{graphicx}

\def\re{{\rm Re}}
\def\im{{\rm Im}}

\def\[{\left [}
\def\]{\right ]}
\def\({\left (}
\def\){\right )}
\def\bl{\bar{\lambda}}

\def\pp{\partial}
{

\def\G{{\cal G}}

\newcommand{\pr}{{\it Phys.\ Rev. }}
\newcommand{\pl}{{\it Phys.\ Lett. }}
\newcommand{\np}{{\it Nucl.\ Phys. }}

\def\L{{\cal L}}

\def\Tev{{\rm TeV}}
\def\Gev{{\rm GeV}}
\newcommand{\be}{\begin{equation}}
\newcommand{\ee}{\end{equation}}
\newcommand{\bea}{\begin{eqnarray}}
\newcommand{\eea}{\end{eqnarray}}

\begin{document}
\begin{titlepage}

      \hfill  LBNL-50425 

      \hfill  UCB-PTH-02/24

      \hfill hep-ph/0205317

\hfill May 2002 \\[.2in]
\begin{center}

{\large \bf PROBING NEW PHYSICS: FROM CHARM TO 
SUPERSTRINGS}\footnote{Talk 
presented at Arkadyfest, University of 
Minnesota, Minneapolis, MN, May 17-18, 2002, to be published in the 
proceedings.}\footnote{This work was 
supported in part by the Director, Office of
Energy Research, Office of High Energy and Nuclear Physics, Division
of High Energy Physics of the U.S. Department of Energy under
Contract DE-AC03-76SF00098 and in part by the National Science
Foundation under grants PHY-0098840 and INT-9910077.}

Mary K. Gaillard \\[.05in]

{\em Department of Physics,University of California, and Theoretical
 Physics Group, 50A-5101, Lawrence Berkeley National Laboratory,
 Berkeley, CA 94720, USA}\\[.2in]
\end{center}

\begin{abstract} Effective theories based on experimental data
provide powerful probes and tests of underlying theories in elementary
particle physics.  Examples within and beyond the Standard Model are
discussed, including a specific model for supersymmetry breaking
within the context of the weakly coupled heterotic string.

\end{abstract}
\end{titlepage}

\newpage
\renewcommand{\thepage}{\roman{page}}
\setcounter{page}{2}
\mbox{ }

\vskip 1in

\begin{center}
{\bf Disclaimer}
\end{center}

\vskip .2in

\begin{scriptsize}
\begin{quotation}
This document was prepared as an account of work sponsored by the United
States Government.  While this document is believed to contain
correct information, neither the United States Government nor any agency
thereof, nor The Regents of the University of California, nor any of their
employees, makes any warranty, express or implied, or assumes any legal
liability or responsibility for the accuracy, completeness, or usefulness
of any information, apparatus, product, or process disclosed, or represents
that its use would not infringe privately owned rights.  Reference herein
to any specific commercial products process, or service by its trade name,
trademark, manufacturer, or otherwise, does not necessarily constitute or
imply its endorsement, recommendation, or favoring by the United States
Government or any agency thereof, or The Regents of the University of
California.  The views and opinions of authors expressed herein do not
necessarily state or reflect those of the United States Government or any
agency thereof, or The Regents of the University of California.
\end{quotation}
\vfill

\end{scriptsize}

\vskip 2in

\begin{center}
\begin{small}
{\it Lawrence Berkeley Laboratory is an equal opportunity employer.}
\end{small}
\end{center}

\newpage
\renewcommand{\thepage}{\arabic{page}}
\setcounter{page}{1}

\section{Introduction}
\setcounter{equation}{0} This talk begins with some historical remarks
about work done during the period of the inception of the Standard
Model, especially areas where my own work overlapped significantly
with that of Arkady Vainshtein. The remainder of the talk will be
devoted to more recent work, in which Arkady's work on supersymmetry
also had an influence.  The common thread through the work of both
periods is the interconnectedness of theory and experiment in
unraveling the elementary structure of nature.

Experiment provides us with effective theories.  That is, the data
tell us how to write down effective Lagrangians with which we can
calculate tree-level S-matrix elements that reproduce the data over
some range of energy and distance scales with reasonable accuracy.
When we try to take an effective theory seriously as a quantum field
theory, we typically encounter difficulties (usually in the form of
infinite amplitudes) that suggest a scale at which new physics must
come into play.  Given a hypothesis for the specifics of the new
physics, detailed studies of the effective lower energy theory can
test its validity.  I will recall examples that contributed to the
construction and study of the Standard Model and review approaches to
the phenomenology of supergravity and superstring theory.  A specific
model for supersymmetry breaking within the context of the weakly
coupled heterotic string will be discussed.

\section{Effective Theories for the Standard Model}
An early example of a successful effective
Lagrangian is Fermi theory.  A series of experiments led theorists
such as Fermi, Gamov and Teller to postulate four fermion couplings to
describe nuclear $\beta$-decay.  Further data revealed the V-A nature
of the couplings as well as a set of flavor selection rules for both
strangeness-changing and strangeness-conserving (semi-)leptonic decays of
hadrons.  This led to the interpretation of the effective
Fermi interaction as arising from the exchange of heavy, electrically
charged vector bosons $W^{\pm}$, with the identification $G_F/\sqrt{2}
= g^2_W/m^2_W$, coupled to bilinear quark currents that are Noether
currents of the strong interaction.  These studies provided early
building blocks both of the GWS model of electroweak interactions and
of the quark model and QCD.

The Fermi theory is nonrenormalizable.  Attempts to treat it as
a quantum theory revealed quadratic divergences which suggested that
loop corrections must be effectively cut off by new physics at a 
scale $\Lambda_F\sim 300$ GeV. Similarly, analyses of high energy
scattering amplitudes revealed a breakdown of tree unitarity
at cm energies above 600 GeV.  These results were the motivation
for the construction of the $p\bar p$ colliders that ultimately
produced the $W,Z$ bosons with masses $\sim 100$ GeV, providing 
the needed cut-off. As it happens, by the time of their discovery, 
the underlying electroweak theory had already been developed and
tested, and the measurement of the weak boson masses provided
spectacular confirmation of that theory.  Once the underlying
theory is known, there is no impediment to treating the low energy
effective Fermi theory as a full quantum theory, provided loop
integration is appropriately cut off at the physical threshold for
new physics, for example
\be G^2_F\Lambda^2_F \sim G^2_Fm^2_W = g_W^4m^2_W.\ee

Another example of an effective theory for the Standard Model is the
low energy chiral Lagrangian for pions.  The approximate
$SU(2)_R\otimes SU(2)_L$ invariance of the strong interactions that
was uncovered in the study of pion couplings was a crucial building
block for the theory of QCD with very light $u,d$ quarks.  The
chiral Lagrangian is nonrenormalizable, but its study at the quantum
level, with appropriate cut-offs related to the scale of confinement,
has contributed important information on quark masses and other
aspects of low energy QCD matrix elements.

In the following I discuss two applications of effective theories
for the Standard Model that both Arkady and I were involved in.

\subsection{The Charm Threshold}

The effective Fermi theory constructed from experimental measurements
of weak (semi-)leptonic decays contained both strangeness-changing
$(\bar u s)$ and strangeness-conserving $(\bar u d)$ quark currents.
At the quantum level this leads to a semi-leptonic
$(\bar d s)(\bar\mu\mu)$ four-quark Fermi coupling with effective
Fermi constant
\be G_{\Delta S} \sim \sin\theta_cG^2_F\Lambda^2_{\Delta S},\ee 
and consistency with experiment requires $\Lambda_{\Delta S}\sim 1$
GeV.  The new threshold was provided by the GIM mechanism~\cite{gim}
whereby the $u$ loop contribution was canceled up to quark mass
effects by the loop contribution from the postulated charm quark $c$,
and suggested a charm quark mass $m_c\sim$ GeV.  The BIM
mechanism~\cite{bim}, using the same new quark to cancel gauge
anomalies in the context of the GWS electroweak theory, strengthened
the case for the existence of charm.

In 1973, Ben Lee and I noticed~\cite{gl} that in the limit of $u$-$c$
mass degeneracy, not only the amplitudes for $K\to\mu\mu$ and
$K^0\leftrightarrow \bar K^0$ vanish, but so do those for processes
like $K\to\gamma\gamma$, that are observed to be unsuppressed.  A
similar obervation was made by Ernest Ma~\cite{ma}. A careful analysis
of these and other $K$-decays in the context of the GWS electroweak
theory revealed that unsuppressed processes of the latter type have
amplitudes $\propto\alpha G_F\ln(m_c/ m_u)$, whereas suppressed
processes of the former type have amplitudes $\propto \alpha G_F(m^2_c
- m^2_u)/m^2_W$. Consistency with data required $m^2_c\ll m^2_c\ll
m^2_W$, further supporting very small $u,d$ masses.  While the
calculation of $K\to\mu\mu$ is theoretically very clean, it turns out
that the leading contribution, enhanced by a factor $\ln(m_W/ m_c)$,
cancels between $W$ and $Z$ exchange diagrams; the result was a rather
weak limit: $m_c\le 9$ GeV. On the other hand, if one was brave enough
to attempt to evaluate the matrix element for $K^0\leftrightarrow \bar
K^0$, a prediction on the order of a GeV for the charmed quark could
be inferred and used to predict branching ratios for other rare kaon
decays that had not yet been observed. Once we had done all this work,
we learned from Bjorken that we had been scooped, at least in part, by
our Russian colleagues Vainshtein and Kriplovich~\cite{vk} who had
considered the same processes and made similar inferences about the
charm quark mass.

Ben and I evaluated the $K^0\leftrightarrow \bar K^0$ matrix element
using a simple factorization ansatz, and, from the observed value of
the $K_L$-$K_S$ mass difference, found $m_c\approx 1.5$ GeV, neglecting
color (QCD was just beginning to emerge at that time as a candidate
theory for the strong interactions).  This low value worried
us, since charm had not been seen in experiments (or so it was
assumed; several hints were indeed in the experimental literature;
see, {\it e.g.,}~\cite{glr}).  So we ``rounded it off'' to 2 GeV,
until a reader~\cite{jm} of the draft insisted that we really found
1.5.  So we caved in, but decided to include a second evaluation using
colored quarks which gave us back the seemingly safer value of 2.  Of
course 1.5 is the right answer even though QCD is correct. At the time
we knew nothing of a third generation, and its appearance muddied the
waters for a while, but since the top quark couplings to lighter
quarks are very weak, its contribution to the neutral kaon mass
difference is insignificant. Analyses based on a $1/N_c$ expansion
suggest~\cite{buras} that one should neglect the color factor, and
lattice QCD calculations support~\cite{sharpe} the na\"{\i}ve
factorization hypothesis to a good approximation.

\subsection{Penguins and the $\Delta I = {1\over2}$ Rule}
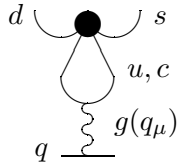
\begin{figure}
\begin{picture}(500,70)(-220,0)
\put(20,50){\circle*{10}}\put(45,50){$s$}\put(-10,50){$d$}
\multiput(10,55)(20,0){2}{\oval(20,20)[b]}
\multiput(20,2)(0,8){3}{\oval(4,4)[l]}\put(35,30){$u,c$}
\multiput(20,6)(0,8){2}{\oval(4,4)[r]}
\put(10,0){\line(1,0){20}}\put(30,10){$g(q_\mu)$}
\put(20,30){\oval(20,20)[b]}\put(0,0){$q$}
\put(20,50){\line(1,-2){10}}\put(20,50){\line(-1,-2){10}}
\end{picture}
\caption{Penguin diagram.\label{fig:peng}}
\end{figure}

While the elementary couplings of the quarks and gluons of QCD become
manifest at high momentum transfer $|q|\gg \Lambda_{QCD}$, at low
energy's we are forced to work with an effective theory of hadrons.
Here I consider weak decays.  Purely leptonic processes can of course
trivially be dealt with using perturbation theory.  Semi-leptonic
processes are also tractable because the hadronic vertex can be
factored out, and the assumption that the hadronic currents are the
Noether currents of flavor/chiral SU(3) is highly predictive.  In
contrast, in nonleptonic decays, the underlying $\bar q q' W$
couplings are completely masked by gluon exchanges across all weak
vertices.  However we learned from Ken Wilson how to find the correct
effective quark Lagrangian at scales much lower than $m_W$: the
operator product expansion~\cite{ken}.  The dominant operator is the
one of lowest dimension, in this case a four quark operator.  Since
gluon exchange conserves helicity, and only left-handed quarks
participate in the charge-changing weak interaction, this is a V-A
Fermi operator, but the I-spin structure is modified with respect to
the original operator.  This is because the eigenstates of the
S-matrix are in fixed representations of color $SU(3)$, and these are
related by Fermi statistics to the final state $ud$ I-spin in the
scattering process $us\to ud$.  The scattering occurs in a $J=L=S=0$
state, which is antisymmetric, so the color and I-spin states must
have the same symmetry.  These are 1) an anti-triplet of $SU(3)_c$
with $I_{ud} = 0$, which is attractive, and 2) a sextet of $SU(3)_c$
with $I_{ud} = 1$, which is repulsive. Since the initial state has
$I={1\over2}$, the former is a pure $\Delta I={1\over2}$ transition
and the second is a mixture of $\Delta I={1\over2}$ and $\Delta
I={3\over2}$.  The explicit calculation~\cite{gl2} gives a mild 
enhancement of the former and suppression of the latter such that 
the ratio of the effective Fermi coupling constants is
$G_F^{1\over2}/G_F^{3\over2}\approx 5$, whereas experiment suggests
that a factor of about 20 is needed, if one evaluates the amplitudes
using simple factorization.  Subsequently, Arkady and his
collaborators pointed out~\cite{svz} that we had forgotten a diagram,
namely the penguin diagram, depicted in Figure~\ref{fig:peng}.
Although our Russian colleagues invented the diagram, its name was
first introduced in a paper~\cite{egnr} (where we also estimated the
$b\bar b$ photoproduction cross section using the SVZ QCD sum
rules~\cite{svz2}) on the properties of the newly discovered B states
after a dart game that ended with John Ellis required to use the word
``penguin'' in his next paper.  (More details on the dart game can be
found in Misha Shifman's contribution to a tribute~\cite{fest} to
Arkady.)  This diagram gives an effective local four-quark operator
because the $q^2$ in the gluon propagator cancels the $q^2$ in the
numerator associated with the color charge radius.  It contributes
only to $\Delta I={1\over2}$ transitions because the penguin's head
has $\Delta I={1\over2}$ ($s\to d$) and its foot has $\Delta I=0$
since gluon exchange conserves flavor. It further has a different spin
structure from the operators analyzed in~\cite{gl2} because, while the
penguin's head has only V-A couplings, its foot is a pure vector
coupling because gluon exchange conserves parity. As a consequence,
after a Fierz transformation so as to express the effective operator
in terms of color singlet quark bilinears, one gets $S,P$ interactions
that can have enhanced matrix elements with respect to the usual $V,A$
couplings.

One is still left with the thorny problem of evaluating hadronic
matrix elements of the effective quark operators.  This involves
techniques such as chiral symmetry and lattice calculations; the
latter have indicated an enhancement of penguin diagrams.  The $\Delta
I={1\over2}$ rule in kaon decay was discussed extensively by
Bijnens~\cite{peng}.  To the extent that the nonrelativistic quark
model is a good approximation for baryons, the $\Delta I={3\over2}$
operator discussed above cannot contribute because it has no overlap
with the baryon wave function~\cite{pw}.  It seems likely to this
observer that the undeservedly accurate ($\sim 2\%$) $\Delta
I={1\over2}$ rule does not have a simple explanation, but results from
the conjunction of a number of QCD effects, all of which point at
least qualitatively in the right direction.

Aside from their contribution to the $\Delta I={1\over2}$ rule, the
SVZ penguin diagrams have played, and continue to play, an essential
role in the analysis of heavy quark decays and CP violation.

\section{Is the Standard Model an Effective Theory?}
Over the last twenty years experiments have continued to verify the
predictions of the Standard Model to great accuracy over a wide range
of energy scales.  Yet the majority of theorists believe that the
Standard Model itself has a limited range of validity.  There are a
variety of reasons for this, one being the large number of arbitrary
parameters.  Here I concentrate on just two, very likely connected,
difficulties: the gauge hierarchy problem and the failure of the
Standard Model to incorporate gravity.  The gauge hierarchy problem
points to a scale of new physics in the TeV region, providing the
motivation for the LHC and an upgraded Tevatron.  The fact that
gravity couples to everything else suggests new physics at a much
higher scale, the Planck scale, that (at least in the ``conservative''
approach taken here) will never be probed directly by experiments;
probing this physics must rely on the interpretation of indirect
effects in the effective low energy theory.  Incorporating gravity in
a fully consistent, predictive and verifiable theory, often referred
to as the ``Theory of Everything'' (ToE) is the holy grail of
elementary particle physics.

\subsection{Bottom Up Approach}
This approach starts from experimental data with the aim of
deciphering what it implies for an underlying, more fundamental
theory.  One
outstanding datum is the observed large gauge hierarchy, {\it i.e.,}
the ratio of the $Z$ mass, characteristic of the scale of electroweak
symmetry breaking, to the reduced Planck scale $m_P$:
$$ m_Z\approx 90\Gev \ll m_P = \sqrt{8\pi\over
G_N}\approx2\times10^{18}\Gev, $$ which can be technically resolved by
supersymmetry (SUSY) (among other conjectures that are by now disfavored
by experiment).  The conjunction of SUSY and general relativity (GR)
inexorably implies supergravity (SUGRA).  The absence of observed SUSY
partners (sparticles) requires broken SUSY in the vacuum, and a more
detailed analysis of the observed particle spectrum constrains the
mechanism of SUSY-breaking in the observable sector: spontaneous
SUSY-breaking is not viable, leaving soft SUSY-breaking as the only
option that preserves the technical SUSY solution to the hierarchy
problem.  This means introducing SUSY-breaking operators of dimension
three or less--such as gauge invariant masses--into the Lagrangian for
the SUSY extension of the Standard Model (SM).  The unattractiveness
of these {\it ad hoc} soft terms strongly suggests that they arise
from spontaneous SUSY breaking in a ``hidden sector'' of the
underlying theory.  Based on the above facts, a number of standard
scenarios have emerged.  These include:
\newline
$\bullet$ Gravity mediated SUSY-breaking, usually understood as
``Minimal SUGRA'' (MSUGRA), with masses of fixed spin particles set
equal at the Planck scale; this scenario is typically characterized
by $ m_{\rm scalars}= m_{0}> m_{\rm gauginos} = m_{1\over2}
\sim m_{\rm gravitino} = m_{3\over2}$ at the weak scale.
\newline
$\bullet$ Anomaly mediated SUSY-breaking~\cite{rs,hit}, in which $m_0
= m_{1\over2}=0$ classically; these models are characterized by $
m_{3\over2} >> m_0 ,\; m_{1\over2},$ and typically $m_0 >
m_{1\over2}$.  An exception is the Randall-Sundrum (RS) ``separable
potential'', constructed~\cite{rs} to mimic SUSY-breaking on a brane
spatially separated from our own in a fifth dimension; in this
scenario $m_0^2 < 0$ and $m_0$ arises first at two loops. More generally,
the scalar masses at one loop depend on the details of Planck-scale
physics~\cite{bgn}.
\newline
$\bullet$ Gauge mediated SUSY uses a hidden sector that has
renormalizable gauge interactions with the SM particles.  These
scenarios are typically characterized by small $m_{1\over2}$.
\begin{figure}
\begin{picture}(350,220)(0,20)
\put(0,36){D = 9}\put(0,116){D = 10}\put(0,196){D = 11}
\multiput(120,40)(160,0){2}{\circle{40}}
\multiput(80,120)(80,0){4}{\circle{40}} \put(200,200){\circle{40}}
\put(68,104){\line(5,-6){40}}\put(260,120){\vector(1,-3){20}}
\put(107,57){\vector(1,-1){1}}\put(293,57){\vector(-1,-1){1}}
\put(332,104){\line(-5,-6){40}}\put(140,120){\vector(-1,-3){20}}
\put(184,188){\vector(-1,-2){24}}\put(216,188){\vector(1,-2){24}}
\put(188,184){\line(0,-1){80}}\put(212,184){\line(0,-1){80}}
\put(188,104){\vector(-1,-1){52}}\put(212,104){\vector(1,-1){52}}
\put(200,40){\vector(-1,0){60}}\put(200,40){\vector(1,0){60}}
\put(262,120){\vector(-1,0){2}}\put(262,136){\vector(-1,0){10}}
\put(304,134){\vector(0,-1){2}}\put(336,134){\vector(0,-1){2}}
\put(262,128){\oval(16,16)[r]}\put(320,134){\oval(32,32)[t]}
\put(222,116){WCHS}\put(304,116){$O(32)$}
\put(298,134){I}\put(340,134){H}\put(194,196){M}
\put(116,36){II}\put(270,36){H/I}\put(72,116){IIB}\put(152,116){IIA}
\put(88,80){\circle{8}}\put(130,90){\circle{8}}\put(220,164){\line(1,0){16}}
\put(312,80){\circle{8}}\put(270,90){\circle{8}}\put(172,164){\circle{8}}
\put(162,78){\oval(8,4)}\put(162,78){\oval(16,8)}
\put(230,78){\oval(4,8)}\put(246,78){\oval(4,8)[r]}
\put(230,82){\line(1,0){16}}\put(230,74){\line(1,0){16}}
\put(180,20){$T\leftrightarrow 1/T$}\put(242,140){$T\leftrightarrow
1/T$} \put(300,156){$S\leftrightarrow 1/S$}
\end{picture}
\caption{M-theory according to John Schwarz.
\label{fig:john}}
\end{figure}
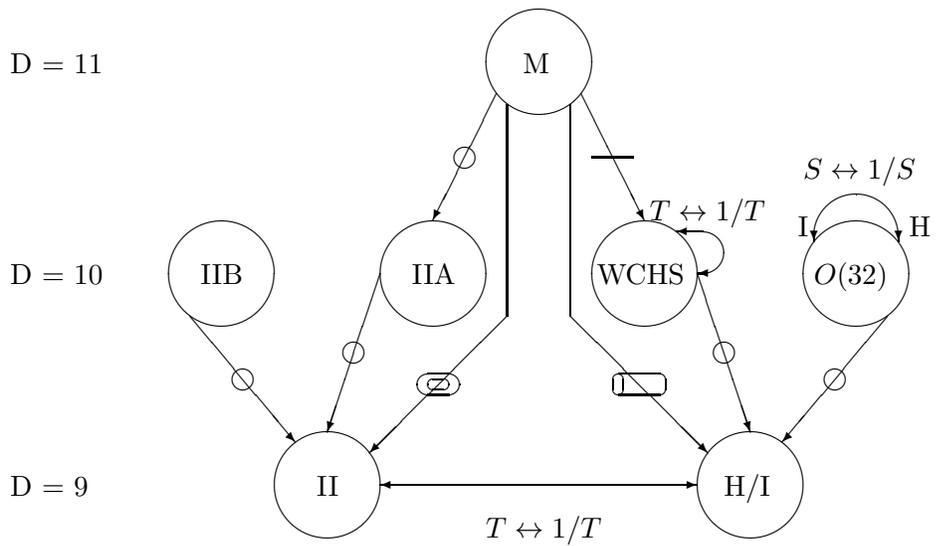
\begin{figure}
\begin{picture}(350,215)(0,40)
\put(150,205){IIA,IIB: D-Branes}\put(120,200){\line(1,0){160}}
\multiput(160,190)(40,0){2}{$\times$}
\put(120,180){\oval(60,40)[l]}\put(280,190){\oval(40,20)[r]}
\put(120,140){\oval(40,40)[r]}\put(280,155){\oval(70,50)[l]}
\put(120,120){\line(-1,0){10}}\put(110,105){\oval(60,30)[l]}
\put(110,80){\oval(30,20)[r]}
\put(110,60){\oval(30,20)[l]}\put(280,110){\oval(80,40)[r]}
\put(280,90){\line(-1,0){10}}\put(270,75){\oval(40,30)[l]}
\put(270,55){\oval(40,10)[r]}\put(110,50){\line(1,0){160}}
\put(230,155){$\times$}\put(255,155){$O(32)_{\rm I}$}
\put(305,110){$\times$}\put(330,110){$O(32)_{\rm H}$}
\put(90,105){$\times$}\put(0,105){11-D SUGRA}\put(190,120){M}
\put(275,53){$\times$}\put(260,40){$E_8\otimes E_8$ WCHS}
\put(105,55){$\times$}\put(0,40){HW theory: {\small (very?) 
large extra dimension(s)}}\end{picture}
\caption{M-theory according to Mike Green.
\label{fig:mike}}\end{figure}
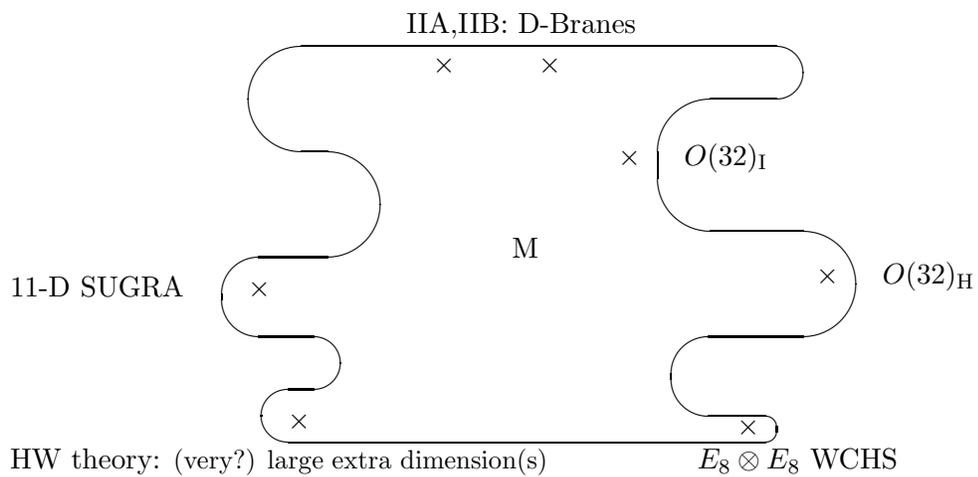

\subsection{Top Down Approach} 
This approach starts from a ToE with the hope of deriving the
Standard Model from it; the current favored candidate is
superstring theory.  The driving motivation is that this is at present
the only known candidate for reconciling GR with quantum
mechanics. Superstring theories are consistent in ten dimensions; 
in recent years it was discovered that all the consistent
superstring theories are related to one another by dualities.  These
are, in my nomenclature: S-duality: $\alpha\to 1/\alpha,$ and
T-duality: $R\to 1/ R,$ where $\alpha$ is the fine
structure constant of the gauge group(s) at the string scale, and
$R$ is a radius of compactification from dimension D to
dimension ${\rm D} -1$.  Figure~\ref{fig:john} shows~\cite{john} how
these dualities relate the various 10-D superstring theories to one
another, and to the currently presumed ToE, M-theory.  Not a lot else
is known about M-theory, except that it lives in 11 dimensions and
involves membranes.  In Figure~\ref{fig:john} the small circles, line,
torus and cylinder represent the relevant compact manifolds in
reducing D by one or two.  The two $O(32)$ theories are S-dual to one
another, while the $E_8\otimes E_8$ weakly coupled heterotic string
theory (WCHS) is perturbatively invariant~\cite{mod} under T-duality.
We will be specifically concentrating on this theory, and T-duality
will play an important role.

Another image of M-theory, the ``puddle diagram'' of
Figure~\ref{fig:mike}, indicates~\cite{mike} that all the known
superstring theories, as well as \mbox{D $=11$} SUGRA, are particular
limits of M-theory.  Currently, there is a lot of activity in type I
and II theories, or more generally in theories with branes.  Similarly
the Ho\v rava-Witten (HW) scenario~\cite{hw} and its inspirations have
received considerable attention.  If one compactifies one dimension of
the 11-D limit of M-theory, one gets the HW scenario with two 10-D
branes, each having an $E_8$ gauge group.  As the radius of this 11th
dimension is shrunk to zero, the WCHS scenario is recovered.  This is
the scenario addressed here, in a marriage of the two approaches that
may serve as an illustrative example of the diversity of possible SUSY
breaking scenarios.

\section{The $E_8\otimes E_8$ Heterotic String}
I first recall the reasons for the original appeal of the weakly
coupled $E_8\otimes E_8$ heterotic string theory~\cite{gross}
compactified on a Calabi-Yau (CY) manifold~\cite{cy} (or a CY-like
orbifold~\cite{orb}).  The zero-slope (infinite string tension) limit
of this superstring theory~\cite{gs} is ten dimensional supergravity
coupled to a supersymmetric Yang-Mills theory with an $E_8\otimes E_8$
gauge group.  To make contact with the real world, six of these ten
dimensions must be unobservable in current experiments;
here they are assumed to be compactified to a size of order
of the reduced Planck length, $10^{-32}$cm.  If the topology of the
extra dimensions were a six-torus, which has a flat geometry, the
8-component spinorial parameters of $N=1$ supergravity in ten
dimensions would appear as the four two-component parameters of $N=4$
supergravity in ten dimensions.  However a Calabi-Yau manifold leaves
only one of these spinors invariant under parallel transport; for this
manifold the group of transformations under parallel transport
(holonomy group) is the $SU(3)$ subgroup of the maximal $SU(4) \cong
SO(6)$ holonomy group of a six dimensional compact space.  This breaks
$N=4$ supersymmetry to $N=1$ in four dimensions.  As is well known,
the only phenomenologically viable supersymmetric theory at low
energies is $N=1$, because it is the only one that admits complex
representations of the gauge group that are needed to describe quarks
and leptons. For this solution, the classical equations of motion
impose the identification of the affine connection of general
coordinate transformations on the compact space (described by three
complex dimensions) with the gauge connection of an $SU(3)$ subgroup
of one of the $E_8$'s: $E_8\ni E_6\otimes SU(3)$, resulting in
$E_6\otimes E_8$ as the gauge group in four dimensions.  Since the
early 1980's, $E_6$ has been considered the largest group that is a
phenomenologically viable candidate for a Grand Unified Theory (GUT)
of the Standard Model.  Hence $E_6$ is identified as the gauge group
of the ``observable sector'', and the additional $E_8$ is attributed
to a ``hidden sector'', that interacts with the former only with
gravitational strength couplings.  Orbifolds, which are flat spaces
except for points of infinite curvature, are more easily studied than
CY manifolds, and orbifold compactifications that closely mimic the CY
compactification described above, and that yield realistic spectra
with just three generations of quarks and leptons, have been
found~\cite{iban}.  In this case the surviving gauge group is
$E_6\otimes\G_o \otimes E_8,\;\G_o\in SU(3)$.  The low energy
effective field theory is determined by the massless spectrum, {\it
i.e.}, the spectrum of states with masses very small compared with the
scales of the string tension and of compactification. Massless particles
have zero triality under an $SU(3)$ which is the diagonal of the
$SU(3)$ holonomy group and the (broken) $SU(3)$ subgroup of one $E_8$.
The ten-dimensional vector fields $A_M,\; M = 0,1,\ldots 9,$ appear in
four dimensions as four-vectors $A_\mu,\;\mu = M = 0,1,\ldots 3$, and
as scalars $A_m,\; m = M-3 = 1,\cdots 6.$ Under the decomposition
$E_8\ni E_6\otimes SU(3)$, the $E_8$ adjoint contains the adjoints of
$E_6$ and $SU(3)$, and the representation ${\bf(27,3)} +
{\bf(\overline{27},\overline{3})}$.  Thus the massless spectrum
includes gauge fields in the adjoint representation of
$E_6\otimes\G_o\otimes E_8$ with zero triality under each $SU(3)$,
and scalar fields in ${\bf 27 + \overline{27}}$ of $E_6$, with
triality $\pm1$ under each $SU(3)$, together with their fermionic
superpartners.  The number of ${\bf 27}$ and ${\bf\overline{27}}$
chiral supermultiplets that are massless depends on the detailed
topology of the compact manifold.  The important point for
phenomenology is the decomposition under $E_6\to SO(10)\to SU(5)$: \be
({\bf 27})_{E_6} = ({\bf 16 + 10 + 1})_{SO(10)} = \({\bf \{\bar{5} +
10 + 1\} + \{5 + \bar{5}\} + 1}\)_{SU(5)}.\ee A ${\bf \overline{5} +
10 + 1}$ contains one generation of quarks and leptons of the Standard
Model, a right-handed neutrino and their scalar superpartners; a ${\bf
5 + \overline{5}}$ contains the two Higgs doublets needed in the
supersymmetric extension of the Standard Model and their fermion
superpartners, as well as color-triplet supermultiplets. Thus all the
states of the Standard Model and its minimal supersymmetric extension
are present.  On the other hand, there are no scalar particles in the
adjoint representation of $E_6$. In conventional models for
grand unification, adjoints (or one or more other representations much
larger than the fundamental one) are needed to break the GUT group to
the Standard Model.  In string theory, this symmetry breaking can be
achieved by the Hosotani, or ``Wilson line'', mechanism~\cite{hos} in
which gauge flux is trapped around ``tubes'' in the
compact manifold, in a manner reminiscent of the Arahonov-Bohm effect.
The vacuum value of the trapped flux $<\int d\ell^m A_m>$ has the same
effect as an adjoint Higgs, evading the difficulties of 
constructing a viable Higgs sector encountered in conventional GUTS.
Wilson lines reduce the gauge group in four dimensions to
\bea &&\G_{obs}\otimes\G_{hid},
\quad\G_{obs}=\G_{SM}\otimes\G'\otimes\G_o,\quad \G_{SM}\otimes\G'\in
E_6, \quad \G_o\in SU(3),\nonumber \\ && \G_{hid}\in E_8,\quad \G_{SM}
= SU(3)_c\otimes SU(2)_L\otimes U(1)_w.
\label{eq:group}\eea

There are many other four dimensional string vacua with different
features. The attractiveness of the above 
picture is that the requirement of $N=1$ SUSY naturally results in a
phenomenologically viable gauge group and particle spectrum. Moreover,
the gauge symmetry can be broken to a product group embedding the
Standard Model without the necessity of introducing large Higgs
representations.  In addition, the $E_8\otimes E_8$ string theory
includes a hidden sector that can provide a viable mechanism for
spontaneous SUSY breaking.  More specifically, if some subgroup $\G_a$
of $\G_{hid}$ is asymptotically free, with a $\beta$-function
coefficient $b_a>b_{SU(3)}$, defined by the renormalization group
equation (RGE)
\be \mu{\pp g_a(\mu)\over\pp\mu} = -{3\over2}b_ag_a^3(\mu) +
O(g_a^5)\label{eq:rge},\ee 
confinement and fermion condensation will
occur at a scale $\Lambda_c\gg\Lambda_{QCD}$, and hidden sector
gaugino condensation $<\bl\lambda>_{\G_a} \ne 0,$ may
induce~\cite{nilles} supersymmetry breaking.  

To discuss supersymmetry breaking in more detail, we need the low
energy spectrum resulting from the ten-dimensional gravity
supermultiplet that consists of the 10-D metric $g_{MN}$, an
antisymmetric tensor $b_{MN}$, the dilaton $\phi$, the gravitino
$\psi_M$ and the dilatino $\chi$.  For the class of CY and orbifold
compactifications described above, the zero-triality
massless bosons in four
dimensions are the 4-D metric $g_{\mu\nu}$, the antisymmetric tensor
$b_{\mu\nu}$, the dilaton $\phi$, and certain components of the
tensors $g_{mn}$ and $b_{mn}$ that form the real and imaginary parts,
respectively, of complex scalars known as moduli.  (More precisely,
the scalar components of the chiral multiplets of the low energy
theory are obtained as functions of the scalars $\phi,g_{mn}$, while
the pseudoscalars $b_{mn}$ form axionic components of these
supermultiplets.)  The number of moduli is related to the number of
particle generations (\# of ${\bf 27}$'s $-$ \# of
${\bf\overline{27}}$'s).  Typically, in a three generation orbifold
model there are three moduli $t_I$; the $vev$'s $<{\rm Re}t_I>$
determine the radii of compactification of the three tori of the
compact space.  In some compactifications there are three other moduli
$u_I$; the $vev$'s $<{\rm Re}u_I>$ determine the ratios of the two
{\it a priori} independent radii of each torus.  These form chiral
multiplets with fermions $\chi^t_I, \chi^u_I$ obtained from components
of $\psi_m$.  The 4-D dilatino $\chi$ forms a chiral multiplet with
with a complex scalar field $s$ whose $vev$ $ <s> = g^{-2} -
i\theta/8\pi^2$ determines the gauge coupling constant and the
$\theta$ parameter of the 4-D Yang-Mills theory.  The ``universal''
axion Im$s$ is obtained by a duality transformation~\cite{wit} from
the antisymmetric tensor $b_{\mu\nu}$: $\pp_\mu{\rm
Im}s\leftrightarrow \epsilon_{\mu\nu\rho\sigma}\pp^\nu
b^{\rho\sigma}.$ Because the dilaton couples to the (observable and
hidden) Yang-Mills sector, gaugino condensation induces~\cite{dine} a
superpotential for the dilaton superfield $S$ (capital Greek and Roman
letters denote chiral superfields, and the corresponding lower case
letters denote their scalar components):
 \be W(S) \propto e^{-S/b_a}.\label{eq:dil}\ee 
The vacuum value $ <W(S)>$ is governed by the condensation scale 
$\left<e^{-S/b_a}\right> = e^{-g^{-2}/b_a}= \Lambda_c$ as determined by
the RGE (\ref{eq:rge}).  If it is nonzero, the gravitino acquires a
mass $m_{3\over2}\propto<W>$, and local supersymmetry is broken.

\section{A Runaway Dilaton?}
The superpotential (\ref{eq:dil}) results in a potential for the
dilaton of the form $ V(s)\propto e^{-2\re s/b_a},$ which has its
minimum at vanishing vacuum energy and vanishing gauge coupling: $<\re
s> \to\infty,\; g^2\to 0$.  This is the notorious runaway dilaton
problem.  The effective potential for $s$ is determined by
anomaly matching~\cite{vy}: $\delta\L_{eff}(s,u) \longleftrightarrow
\delta\L_{hid}({\rm gauge}),$ where $u, \;\left<u\right> = \left<
\bl\lambda\right>_{\G_a},$ is the lightest scalar bound state of the
strongly interacting, confined gauge sector.  Just as in QCD, the
effective low energy theory of bound states must reflect both the
symmetries and the quantum anomalies of the underlying Yang-Mills
theory.  It turns out that the effective quantum field theory (QFT) is
anomalous under T-duality.  Since this is an exact symmetry of
heterotic string perturbation theory, it means that the effective QFT
is incomplete.  This is cured by including model dependent string-loop
threshold corrections~\cite{thresh} and a ``Green-Schwarz''
(GS) counter-term~\cite{gsterm}, analogous to a similar anomaly
canceling mechanism in 10-D SUGRA~\cite{gs}.  This introduces
dilaton-moduli mixing, and the gauge coupling constant is now
identified as $g^2= 2\left<\ell\right>,\;\ell^{-1} = 2\re s -
b{\sum_I} \ln(2\re t_I), $ where $b\le b_{E_8} = 30/8\pi^2 $ is the
coefficient of the GS term.  This term introduces a second runaway
direction at strong coupling: $V \to - \infty$ for
$g^2\to\infty$. The small coupling behavior is unaffected, but the
potential becomes negative for $\alpha = \ell/2\pi > .57$. This is the
strong coupling regime, and nonperturbative string effects cannot be
neglected; they are expected~\cite{shenk} to modify the K\"ahler
potential for the dilaton, and therefore the potential $V(\ell,u)$.
It has been shown~\cite{us,casas} that these contributions can indeed
stabilize the dilaton.

In order to carry out the above program, one first needs to know the
quantum corrections to the unconfined ({\it i.e.,} at scales greater
than $\Lambda_c$) Yang-Mills Lagrangian for the hidden sector, in
order to do the correct anomaly matching.  This is where we again
encounter Arkady's work.  Just as we can correctly treat the effective
Fermi Lagrangian as a quantum theory only if we incorporate the
physical cut-off $m_W$, we must include a similar physical cut-off
when encountering divergences in the effective SUGRA theory from
superstrings.  One way to do this is using Pauli-Villars regulators
with couplings that respect SUSY~\cite{pv}.  The required
cut-offs~\cite{pv,gt} for regulating the coefficient of the Yang-Mills
terms are (neglecting string nonperturbative corrections to the
K\"ahler potential)
\be \Lambda_A = \Lambda(\Lambda/g)^{2(1 - 3q_A)}, \quad
\Lambda_g = \Lambda g^{-{2\over3}},\label{cutoffs}\ee
for matter and gauge loops, respectively, where $\Lambda = R^{-1}$ is
the inverse radius of compactification, and $q_A$ is the average
modular weight for the chiral supermultiplet $\Phi^A$. For
``untwisted'' matter $q_A = {1\over3}$, giving the intuitive value
$\Lambda_U = \Lambda = 1/R$, while the gauge-loop cut-off contains the
``two-loop'' factor $g^{-{2\over3}}$ which assures that the anomaly is
a chiral superfield.  Matching~\cite{gt} the field theory result with
string loop calculations~\cite{thresh} determines the GS term for
$\Phi^A=0$, and one obtaines for the Wilson coefficient of the Yang-Mills 
operator
\be (g^{-2}_a)_W = {2\ell}^{-1} -{1\over16\pi^2}(C^a_G-C^a_M)\ln\ell
-{1\over8\pi^2}\sum_AC^a_A\ln (1-p_A\ell)\, + O(|\phi^A|^2) +
\Delta_a(t), \label{deltal}\ee
where $\Delta_a(t)$ is a string threshold correction that is present
in some orbifold models, and $p_A$ is the coupling of $|\Phi^A|^2$ to
the GS term.  Neglecting $\Delta_a$, the RGE invariant
(\ref{deltal}) can be identified with the general RGE invariant in
SUSY Yang-Mills theories found by Shifman and Vainshtein~\cite{sv}
\be g_a^{-2}(\mu)-{1\over16\pi^2}
(3C_G^a-C_M^a)\ln\mu^2+{C_G^a8\pi^2}\ln g^2_a(\mu)
+{1\over8\pi^2}\sum_AC_A^a\ln Z_A^a(\mu)\, ,
\label{rge}\ee
where $Z_A^a$ are the renormalization factors for the matter fields,
provided we identify
\be g_a(\mu_s ) = 2\ell = (\mu_s/M_P)^2,\quad Z_A^a(\mu_s)=(1-p_A
\ell)^{-1}.\label{id}\ee
The same boundary condition on the couplings was found in~\cite{kl}
where different regularization procedures were used.  (The inclusion
of string nonperturbative effects can modify these results; their
effects were found to be negligible in the model of~\cite{us}).

\section{A Condensation Model for SUSY Breaking}
In this section I discuss features of an explicit model~\cite{us}
 based on affine level one orbifolds with three untwisted moduli $T_I$
 and a gauge group of the form (\ref{eq:group}).  Retaining just one
 or two terms of the suggested parameterizations~\cite{shenk} of the
 nonperturbative string corrections:
 $a_n\ell^{-n/2}e^{-c_n/\sqrt{\ell}}$ or $a_n\ell^{-n}e^{-c_n/\ell},$
 the potential can be made positive definite everywhere and the
 parameters can be chosen to fit two data points: the coupling
 constant $g^2\approx 1/2$ and the cosmological constant
 $\Lambda_{cos} \simeq 0$.  This is fine tuning, but it can be done
 with reasonable (order 1) values for the parameters $c_n, a_n$.  If
 there are several condensates with different $\beta$-functions, the
 potential is dominated by the condensate with the largest
 $\beta$-function coefficient $b_+$, and the result is essentially the
 same as in the single condensate case, except that a small mass is
 generated for the axion Im$s$.  In this model, mass hierarchies arise
 from the presence of $\beta$-function coefficients; these have
 interesting implications for both cosmology and the spectrum of
 sparticles.

\subsection{Modular Cosmology}

The masses of the dilaton $d=\re s$ and the
complex $t$-moduli are related to the gravitino mass by~\cite{us}
\be m_d \sim {1\over b_+^2} m_{\tilde G}, \quad m_{t_I} \approx
{2\pi\over3}{(b-b_+)\over(1+b<\ell>)}m_{\tilde G}.
\label{eq:modmass} \ee
Taking $b = b_{E_8} \approx .38 \approx 10b,$ gives a
hierarchy of order $m_{3\over2}\sim 10^{-15}m_{Pl}\sim 10^3$ GeV and
$m_{t_I}\approx 20 m_{3\over2}\approx20$ TeV, $m_d\sim
10^3m_{2\over3}\sim 10^6$ GeV, which is sufficient to evade the late
moduli decay problem~\cite{modprob} in nucleosynthesis.

If there is just one hidden sector condensate, the axion $a = \im s$
is massless up to QCD-induced effects:
$m_a\sim(\Lambda_{QCD}/\Lambda_c)^{3\over2}m_{3\over2}\sim10^{-9}$ eV,
and it is the natural candidate for the Peccei-Quinn axion. Because of
string nonperturbative corrections to its gauge kinetic term, the
decay constant $f_a$ of the canonically normalized axion is reduced
with respect to the standard result by a factor $b_+ \ell^2
\sqrt{6}\approx 1/50$ if $b_+\approx .1b_{E_8}$, which may be
sufficiently small to satisfy the (looser) constraints on $f_a$ when
moduli are present~\cite{bd}.

\subsection{Sparticle Spectrum}

\begin{figure}
\includegraphics{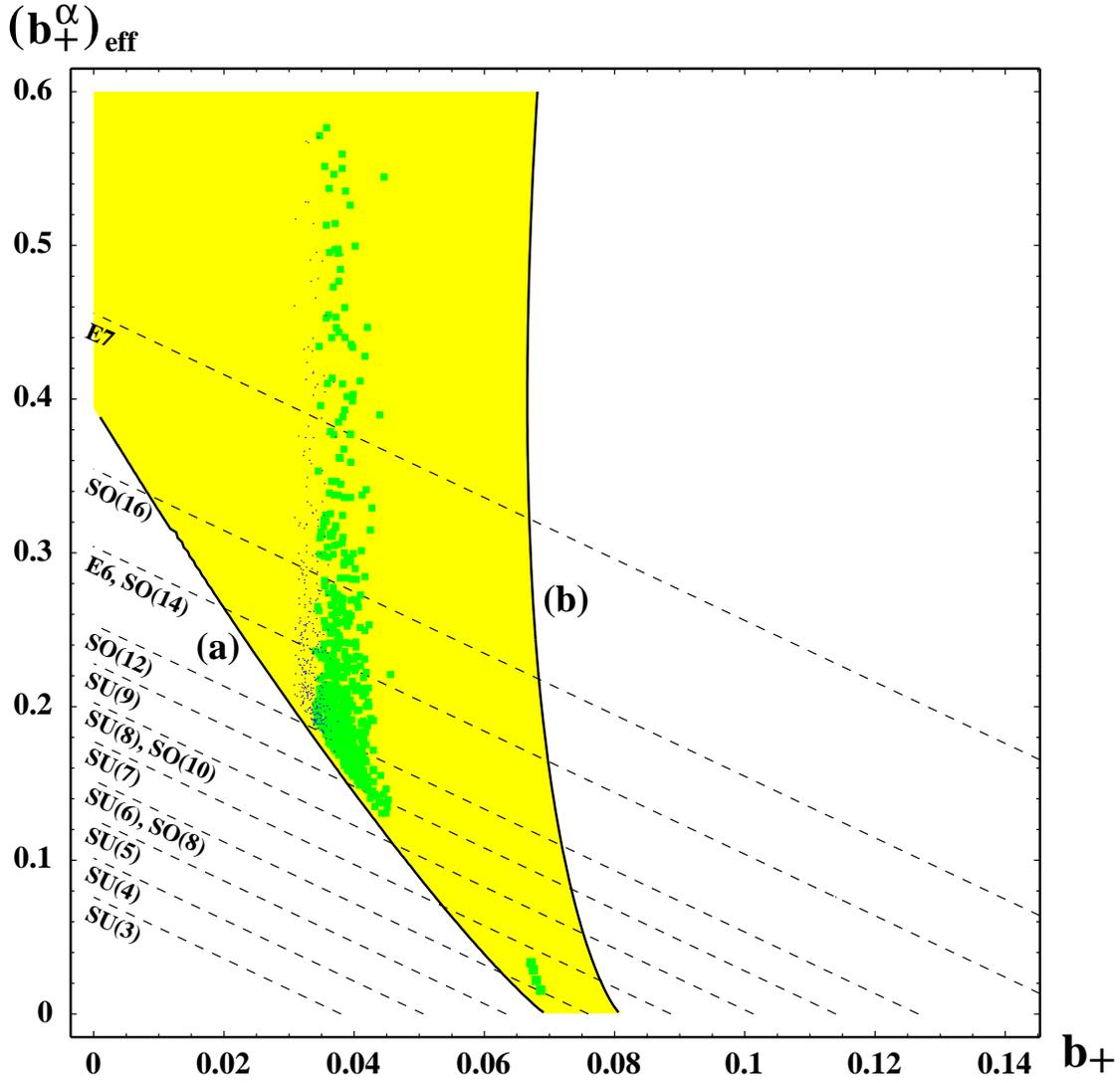}
\caption{Viable hidden sector gauge groups for scenario A of the
condensation model. The swath bounded by lines (a) and (b) is the
region defined by $.1<m_{3\over2}/\Tev,\;\lambda_c <10$, where
$\lambda_c$ is a condensate superpotential coupling constant.  The
fine points correspond to $.1\le\Omega_dh^2\le.3$, and the course
points to $.3<\Omega_dh^2\le1$.
\label{fig:abn}}
\end{figure}

In contrast to an enhancement of the dilaton and moduli masses, there
is a suppression of gaugino masses: $m_{1\over2} \approx
b_+m_{3\over2}$, as evaluated at the scale $\Lambda_c$ in the tree
approximation.  As a consequence quantum corrections can be important;
for example there is an anomaly-like scenario in some regions of the
$(b_+,b_+^\alpha)$ parameter space, where $b_+^\alpha$ is the hidden
matter contribution to $b_+.$ If the gauge group for the dominant
condensate (largest $b_a$) is not $E_8$, the moduli $t_I$ are
stabilized at self-dual points through their couplings to twisted
sector matter and/or moduli-dependent string threshold corrections,
and their auxiliary fields vanish in the vacuum. Thus SUSY-breaking is
dilaton mediated, avoiding a potentially dangerous source of flavor
changing neutral currents (FCNC).  These results hold up to the
unknown couplings $p_A$ in (\ref{deltal}): at the scale $\Lambda_c$ $
m_{0A} = m_{3\over2}$ if $p_A = 0$, while $m_{0A}=
{1\over2}m_{t_I}\approx 10 m_{3\over2}$ if the scalars couple with the
same strength as the T-moduli: $ p_A = b$.  In addition, if $p_A = b$
for some gauge-charged chiral fields, there are enhanced loop
corrections to gaugino masses~\cite{gnw}.  Four sample scenarios were
studied~\cite{gn}: A) $p_A=0$, B) $p_A=b$ for the superpartners of the
first two generations of SM particles and $p_A=0$ for the third, C)
$p_A=b$, and D) $p_A=0$ for the Higgs particles and $p_A=b$ otherwise.
Imposing constraints from experiments and the correct electroweak
symmetry-breaking vacuum rules out scenarios B and C.  Scenario A is
viable for $1.65< \tan\beta<4.5$, and scenario D is viable for all
values of $\tan\beta$, which is the ratio of Higgs $vev$'s in the
supersymmetric extension of the SM.  The viable range of
$(b_+,b_+^\alpha)$ parameter space is shown~\cite{abn} in
Figure~\ref{fig:abn} for $g^2={1\over2}$.  The dashed lines represent
the possible dominant condensing hidden gauge groups $\G_+\in E_8$
with chiral matter in the coset space $E_8/\G_{hid}$.

\subsection{Flat Directions in the Early Universe}
Many successful cosmological scenarios--such as an epoch of
inflation--require flat directions in the potential.  A promising
scenario for baryogenesis suggested~\cite{ad} by Affleck and Dine
(AD) requires in particular flat directions during inflation
in sparticle field space: $<\tilde q>, <\tilde\ell>\ne0$, where
$\tilde f$ denotes the superpartner of the fermion $f$.  While flat
directions are common in SUSY theories, they are generally
lifted~\cite{drt} in the early universe by SUGRA couplings to the
potential that drives inflation.  This problem is evaded~\cite{gmo} in
models with a ``no-scale'' structure, such as the classical potential
for the untwisted sector of orbifold compactifications. Although the
GS term breaks the no-scale property of the theory, quasi-flat
directions can still be found. An explicit model~\cite{lyth} for
inflation based on the effective theory described above allows
dilaton stabilization within its domain of
attraction with one or more moduli stabilized at the vacuum value
 $t_I=e^{i\pi/6}$. One of the moduli may be the inflaton.
The moduli masses (\ref{eq:modmass}) are sufficiently large to evade the
late moduli decay problem in nucleosynthesis, but unlike the dilaton,
they are insufficient to avoid a large relic LSP density without
violation~\cite{lsp} of R-parity (a quantum number that distinguishes
SM particles from their superpartners). If R-parity is conserved, this
problem can be evaded if the moduli are stabilized at or near their
vacuum values--or for a modulus that is itself the inflaton.  It is
possible that the requirement that the remaining moduli be in the
domain of attraction is sufficient to avoid the problem altogether.
For example, if $\im t_I = 0$, the domain of attraction near $t_I = 1$
is rather limited: $0.6<{\rm Re}t_I<1.6$, and the entropy produced by
dilaton decay with an initial value in this range might be less than
commonly assumed.  The dilaton decay to its true ground state may
provide~\cite{cgmo} partial baryon number dilution, which is generally
needed for a viable AD scenario.

\subsection{Relic Density of the Lightest SUSY Particle (LSP)}
Two pertinent questions for SUSY cosmology are:
\newline
$\bullet$ Does the LSP overclose the Universe?
\newline
$\bullet$ Can the LSP be dark matter?
\newline
The window for LSP dark matter in
the much-studied MSUGRA scenario~\cite{efo}, has become ever more tiny
as the Higgs mass limit has increased; in fact there is not much
parameter space in which the LSP does not overclose the universe.  The
ratios of electroweak sparticle masses at the Plank scale determine the
composition of the LSP (which must be neutral) in terms of the Bino
(superpartner of the SM $U(1)$ gauge boson), the Wino (superpartner of
the SM $SU(2)$ gauge boson), and the higgsino (superpartner of the
Higgs boson).  The MSUGRA assumption of equal gaugino masses at the
Planck scale leads to a Bino LSP with rather weak couplings, resulting
in little annihilation and hence the tendency to overclose the
universe, except in a narrow range of parameter space where the LSP is
nearly degenerate with the next to lightest sparticle (in this case a
stau $\tilde\tau$), allowing significant coannihilation.  Relaxing
this assumption~\cite{abn} it was found that a predominantly Bino LSP
with a small admixture of Wino can provide the observed density fraction
$\Omega_d$ of dark matter.  In the model of~\cite{us}, this occurs in
the region indicated by fine points in Figure~\ref{fig:abn}.  In this
model the deviation from the MSUGRA scenario is due to the importance
of loop corrections to small tree-level gaugino masses; in addition to
a small Wino component in the LSP, its near degeneracy in
mass with the lightest charged gaugino enhances
coannihilation.  For larger $b_+$ the LSP becomes pure Bino as in
MSUGRA, and for smaller values it becomes Wino-dominated as in
anomaly-mediated models which are cosmologically safe, but do not
provide LSP dark matter, because Wino annihilation is too fast.

\subsection{Realistic Orbifold Models?}
Orbifold compactifications with the Wilson line/Hosotani mechanism
needed to break $E_6$ to the SM gauge group generally have $b_+\le
b\le b_{E_8}$.  An example is a model~\cite{iban} with hidden gauge
group $O(10)$ and $b_+= b = b_{O(10)}$. It is clear from
(\ref{eq:modmass}) that this would lead to disastrous modular
cosmology, since the $t$-moduli are massless.  Moreover, in typical
orbifold compactifications, the gauge group $\G_{obs}\otimes\G_{hid} $
obtained at the string scale has no asymptotically free subgroup that
could condense to trigger SUSY-breaking. However in many
compactifications with realistic particle spectra~\cite{iban}, the
effective field theory has an anomalous $U(1)$ gauge subgroup, which
is not anomalous at the string theory level.  The anomaly is
canceled~\cite{dsw} by a GS counterterm, similar to the GS term
introduced above to cancel the modular anomaly.  This results in a
D-term that forces some otherwise flat direction in scalar field space
to be lifted, inducing scalar $vev$'s that further break the gauge
symmetry and give masses of order $\Lambda_D$ to some chiral
multiplets, so that the $\beta$-function of some of the surviving
gauge subgroups may be negative below the scale $\Lambda_D$, typically
an order of magnitude below the string scale.  The presence of such a
D-term was explicitly invoked in the above-mentioned inflationary
model~\cite{lyth}. Its incorporation into the effective condensation
potential is under study~\cite{gg}.

There is a large vacuum degeneracy associated with the D-term induced
breaking of the anomalous $U(1)$, resulting in many massless
``D-moduli'' that have the potential for a yet more disastrous modular
cosmology~\cite{joel}. However preliminary results indicate that the
D-moduli couplings to matter condensates lift the degeneracy to give
cosmologically safe D-moduli masses.  Although the D-term modifies the
potential for the dilaton, one still {\mbox obtains} moduli stabilized at
self-dual points giving FCNC-free dilaton dominated SUSY-breaking, an
enhanced dilaton mass $m_d$ and a suppressed axion coupling $f_d$.  An
enhancement of the ratio $m_{t_I}/m_{3\over2}$ can result from
couplings to condensates of $U(1)$-charged D-moduli, that also carry
T-modular weights.  However, the requirement of a viable scalar/gaugino
mass ratio may impose severe restrictions on the details of the effective
theory.

\section{Lessons}
History has taught us that high energy
physics can be successfully studied using lower energy effective
theories. Hopefully this will be the case for string theory.  In
particular I have argued that
\newline
$\bullet$ Quantitative studies with predictions for observable phenomena
are possible within the context of the WCHS.
\newline
$\bullet$ Experiments can place restrictions on the underlying theory,
such as the hidden sector spectrum and the couplings and
modular weights of D-moduli when an anomalous $U(1)$ is present.
Data can also inform us about Plank scale physics through
matter couplings to the GS term and one-loop corrections to the soft
SUSY-breaking scalar potential.

Finally, the SUSY-breaking scenario presented here illustrates the
need for sparticle searches to avoid restrictive assumptions based on
``standard scenarios'' that may be misleading in the absence of 
concrete models.

\section*{Acknowledgments}
I am grateful to my many collaborators.  This work was supported in
part by the Director, Office of Energy Research, Office of High Energy
and Nuclear Physics, Division of High Energy Physics of the
U.S. Department of Energy under Contract DE-AC03-76SF00098 and in part
by the National Science Foundation under grants PHY-95-14797 and
INT-9910077.


\begin{thebibliography}{99}
\bibitem{mike} M. Green, a seminar; my apologies if the puddle diagram
was invented by someone else.
\end{thebibliography}
\end{document}